\documentclass[twocolumn,twoside,slac_two]{revtex4}
\usepackage{graphicx}
\usepackage{fancyhdr}
\begin{document}

\title{\boldmath$B^0_s$ Decays and $B$ Leptonic Decays}

%

\author{R. Van Kooten}
\affiliation{Indiana University, Physics Dept., Swain West 117,
Bloomington, IN 47405, U.S.A.}

\begin{abstract}
Branching fractions of $B^0_s$ decays 
into specific $CP$ eigenstates are presented,
and these and other results are combined in world averages to
evaluate implications on the width difference between mass or
$CP$ eigenstates, $\Delta\Gamma_s$.  New results on purely leptonic
decays of $B$ hadrons 
from the Tevatron and $B$ factories are also presented.
\end{abstract}

\maketitle

\thispagestyle{fancy}


\section{Overview}
Many results from the $B^0_s$ system from the conference
were presented elsewhere. First two-sided limits on the
mass difference of eigenstates or particle-antiparticle oscillation 
frequency $\Delta m_s$ were presented by D\O~\cite{d0buchholz} and
a first measurement by CDF~\cite{bslife}.  Another parameter
describing the $B^0_s$ system is the width difference between
eigenstates, $\Delta \Gamma_s$. Following an introduction to the 
theory of $\Delta \Gamma_s$, remaining $B^0_s$ branching
fraction results are described and combined with measurements
of the $B^0_s$ lifetime and $\Delta \Gamma_s$
from other conference contributions~\cite{bskklife, bslife} to
form a world average of $\Delta \Gamma_s$ to compare with 
Standard Model (SM) predictions.

A very different topic of $B$ leptonic decays is then covered.
These decays, such as $B^0_s \rightarrow \ell^+ \ell^-$ are
very sensitive to new physics.  Others such as the first observation
of 
$B^+ \rightarrow \tau^+ \nu_{\tau}$\footnote{Charge conjugate
states are included implicitly throughout.}
are  sensitive both to new
physics and to the value of the  CKM matrix element $V_{ub}$.

\section{Theory of \boldmath$B^0_s$ System 
and \boldmath$\Delta \Gamma_s$}

The phenomenon of particle-antiparticle mixing and
oscillations of neutral mesons is both a
fascinating quantum mechanical system,  
and a sensitive probe of 
the flavor sector of the SM, 
parametrized by the CKM matrix. In particular,
for the $B^0_s$ system, we want to probe {\em all} parts of the 
matrix evolution equation:
\begin{equation}
i \frac{d}{dt} 
\left( \begin{array}{c} B^0_s \\ \bar{B}^0_s
       \end{array}  \right)
       =
\left( \begin{array}{cc}
       M - \frac{i \Gamma}{2} & M_{12} - \frac{i \Gamma_{12}}{2} \\
       M_{12}^* - \frac{i \Gamma_{12}^*}{2} & M - \frac{i \Gamma}{2}
       \end{array} \right)
\left( \begin{array}{c} B^0_s \\ \bar{B}^0_s
              \end{array}  \right).
\end{equation}
In the SM, $B^0_s$-$\bar{B}^0_s$ mixing is caused by
flavor-changing weak interaction box diagrams that 
induce non-zero off-diagonal elements in the above~\cite{Bphysrun2}.
The mass eigenstates, defined as the eigenvectors of the above matrix,
are different from the flavor eigenstates, with a 
heavy and light mass eigenstate, respectively:
\begin{equation}
|B_{sH} \rangle = p | B^0_s \rangle + q | \bar{B}^0_s \rangle;
\thinspace \thinspace \thinspace
|B_{sL} \rangle = p | B^0_s \rangle - q | \bar{B}^0_s \rangle,
\end{equation}
with $|p|^2 + |q|^2 = 1$. If $CP$ is conserved in mixing in the
$B^0_s$ system, then $|q| = |p| = 1/\sqrt{2}$, and 
\begin{equation}
| B_{sH} \rangle = |B^{CP{\mathrm{-odd}}} \rangle;
\thinspace \thinspace \thinspace
| B_{sL} \rangle = |B^{CP{\mathrm{-even}}} \rangle,
\end{equation}
i.e., $B^{CP{\mathrm{-odd}}}$ is defined as the
$B^0_s$ state that does not decay to $D_s^+ D_s^-$.
Matrix elements can be extracted experimentally by
measuring a mass and width
difference between eigenstates:
\begin{eqnarray}
\Delta m_s & =  M_H - M_L & \approx 2 |M_{12}|; \nonumber \\
\Delta \Gamma_s & =  \Gamma_L - \Gamma_H & \approx 2 |\Gamma_{12}| \cos\phi.
\end{eqnarray}
Note the sign convention for $\Delta \Gamma_s$ compared to
$\Delta m_s$. In this convention, the SM prediction 
for $\Delta \Gamma_s$ is positive.  The phase angle $\phi$ is
expected to be small in the SM, $\phi \approx 0.3^{\circ}$, and
$\cos\phi$ is often assumed to be unity in $\Delta \Gamma_s$ 
measurements.
Finally, an average width is defined as 
$\Gamma_s = (\Gamma_L + \Gamma_H)/2$.  Note that the 
measured lifetime of the $B^0_s$ will depend on the mix of $CP$
eigenstates involved in its decay.  A more fundamental lifetime
based on the average width is defined as 
$\bar\tau = 1/\Gamma_s$, with lifetimes of the light and
heavy mass eigenstates defined as: $\tau_L = 1/\Gamma_L$ and 
$\tau_H = 1/\Gamma_H$.

\begin{figure}[htb]
\centering
\includegraphics[width=80mm]{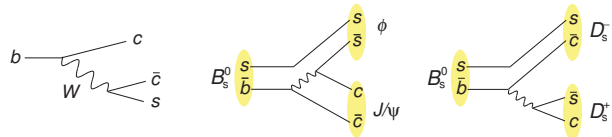}
\caption{Example $B^0_s$ decays giving rise to a 
non-zero $\Gamma_{12}$.} \label{fig:bsdecays}
\end{figure}

The parameter $\Gamma_{12}$ is dominated by the decay path 
$b \rightarrow c \bar{c}s$ in decays into final states
common to both $B^0_s$ $(\bar{b}s)$ and $\bar{B}^0_s$ $(b\bar{s})$.
Examples of such decays are 
$B^0_s \rightarrow J/\psi \phi$ and 
$B^0_s \rightarrow D^{(*)+}_s D^{(*)-}_s$, as shown in
Fig.~\ref{fig:bsdecays}.

The analogous decay diagram for a width
difference in the $B^0_d$ system substitutes a $d$
quark for the $s$ quark. This decay is Cabibbo suppressed,
hence $\Delta \Gamma_d$ is negligible.
In the case of $\Delta \Gamma_s$, decays into
$CP$-even final states increase the value of $\Delta \Gamma_s$,
while decays into $CP$-odd final states decrease it.

\section{Measurements of 
\boldmath$Br(B^0_s \rightarrow D_s^{(*)+} D_s^{(*)-})$}

The decay $B^0_s \rightarrow D_s^{+} D_s^{-}$ is into
a final state that is purely $CP$ even. 
Under various theoretical assumptions~\cite{Aleksan:1993qp}, the
inclusive decay into these ground states plus the excited states
$B^0_s \rightarrow D_s^{(*)+} D_s^{(*)-}$ is also $CP$ even
to within 5\%. This conclusion will likely need re-examination
due to the restrictive assumptions taken; however, proceeding
with this assumption, measurements of this branching fraction
can be used to extract $\Delta \Gamma_s$ (for the phase
angle $\phi = 0$):
\begin{equation}
\frac{\Delta \Gamma_s}{\Gamma_s} \approx
\frac{2 Br(B^0_s \rightarrow D_s^{(*)+} D_s^{(*)-})}
{1 - Br(B^0_s \rightarrow D_s^{(*)+} D_s^{(*)-})/2}.
\label{dGsbr}
\end{equation}
In the above and the following, since flavor tagging
is {\em not} used, branching fractions include 
$\bar{B}^0_s$ in the initial state and are properly
averaged.
There are new measurements for the branching fraction for this
decay channel from CDF, D\O, and Belle.
Only one measurement has been previously published~\cite{aleph_bsbr}
by ALEPH from a study of correlated $\phi\phi$ production
in $Z^0$ decays.

\subsection{CDF Measurement of 
\boldmath$Br(B^0_s \rightarrow D_s^+ D_s^-)$}

\begin{figure}[htb]
\centering
\includegraphics[width=80mm]{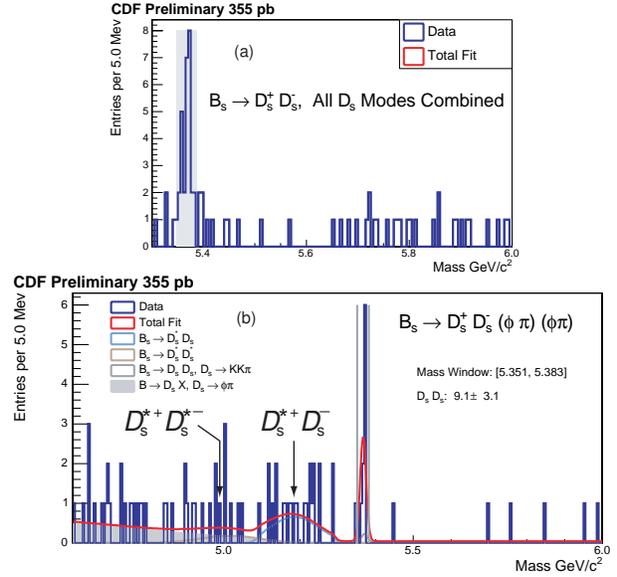}
\caption{
(a) CDF Combined $B^0_s \rightarrow D_s^+ D_s^-$ invariant
mass plot. Grey band shows the 40 MeV wide signal region.
(b) $B^0_s \rightarrow D_s^+ D_s^- (\phi \pi^+)(\phi \pi^-)$
wide range data fit.  $D^{(*)}_s D_s$ and 
$D^{(*)}_s D^{(*)}_s$ modes fitted with templates are also
shown.
} \label{fig:cdfbsbr}
\end{figure}

The CDF Collaboration uses 355~pb$^{-1}$ of data
to fully reconstruct the decay
$B^0_s \rightarrow D^+_s D^-_s$ for the first time~\cite{cdfbsbr}, 
where the $D_s$ is reconstructed via fully hadronic
decays $D_s \rightarrow \phi(\rightarrow K^+K^-) \pi$, $3\pi$, or
$K^*K$.  The rate is then normalized to 
the larger signal of $B^0_s \rightarrow D_s^+ D^-$, where
$D^- \rightarrow K \pi \pi$.  
In addition, many more hadronic channels with similar 
topology (decays into two secondary resonances with
three tracks each) are studied in detail with larger statistics.
Combining all $D_s$ modes, CDF obtains a clean signal
of $23.5 \pm 5.5$ candidates with negligible background as
shown in Fig.~\ref{fig:cdfbsbr}. 

Using the PDG~\cite{pdg} value
of $Br(D_s \rightarrow \phi \pi)$, CDF obtains as
a preliminary result:

\begin{eqnarray}
\frac{Br(B^0_s \rightarrow D^+_s D^-_s)}
{Br(B^0_s \rightarrow D^+_s D^-)}
= & 1.67 \pm 0.41 \thinspace {\mathrm{(stat.)}} 
       \pm    0.12 \thinspace {\mathrm{(syst.)}} \nonumber \\
  &	\pm   0.24 \thinspace (f_s/f_d)
\pm	   0.39 \thinspace (Br_{\phi \pi}).
\end{eqnarray}

Work continues by CDF 
to use this ratio of branching fractions to
extract $\Delta\Gamma_s$.
Hints for the other modes into excited states exist,
and there are good prospects with 1~fb$^{-1}$ of data.

\subsection{D\O\ Measurement of
\boldmath$Br(B^0_s \rightarrow D_s^{(*)+} D_s^{(*)-})$}

\begin{figure}[htb]
\centering
\includegraphics[width=60mm]{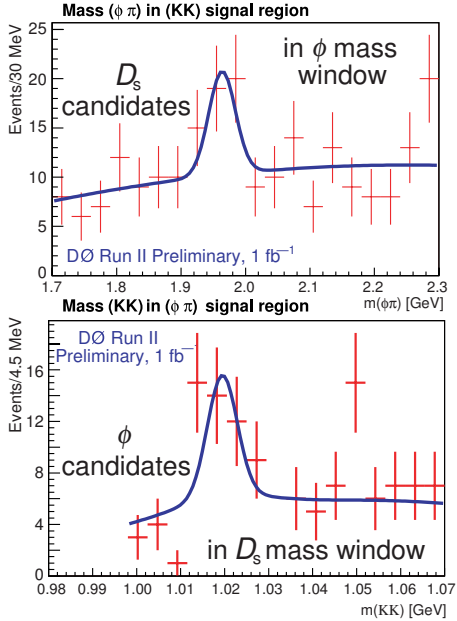}
\caption{
Top: D\O\ $M(\phi \pi)$ invariant mass for candidates
within a second $\phi$ mass window indicating
$D_s$ decays; bottom:
$M(K^+K^-)$ invariant mass for candidates within
a $D_s$ mass window showing $\phi$ decays.
} \label{fig:d0bsbr}
\end{figure}

The D\O\ Collaboration has made a measurement~\cite{d0bsbr} of
the inclusive branching fraction
$Br(B^0_s \rightarrow D_s^{(*)+} D_s^{(*)-})$,
where one $D_s^{(*)}$ is reconstructed in the
hadronic mode $\phi (\rightarrow K^+ K^-) \pi$, and 
the other semileptonically into $\phi \mu \nu$.
Normalization is made to large statistics decay 
mode $B^0_s \rightarrow D_s^{(*)+} \mu^- \nu$,
i.e., measuring:
\begin{equation}
R = \frac{Br(B^0_s \rightarrow D_s^{(*)+} D_s^{(*)-})
\cdot Br(D_s \rightarrow \phi \mu \nu)}
{Br(B^0_s \rightarrow D_s^{(*)-} \mu^+ \nu)}, 
\label{d0ratio}
\end{equation}
where many systematic uncertainties cancel the ratio.
A sample of 15.2k candidates 
of $B^0_s \rightarrow D_s^{-} \mu^+ \nu X$
in approximately 1~fb$^{-1}$ of data 
is first isolated where $D_s^- \rightarrow \phi \pi^-$
is observed in association with a 
close-by $\mu^+$. 
An additional $\phi$ is then searched for in this sample.
Correlated production of excess $\phi$ mesons when
examining a $D_s$ invariant mass window, and excess 
$D_s$ mesons when examining a $\phi$ mass window is
observed as shown in Fig.~\ref{fig:d0bsbr}.

A simultaneous unbinned likelihood fit to
these distributions, and also to mass sidebands to
estimate backgrounds, finds a total number of 
$\mu \phi D_s$ candidates of $19.3 \pm 7.8$.
Small estimated contributions
from 
$B \rightarrow 
D_s^{(*)+} D_s^{(*)-} K X$, 
$B^0_s \rightarrow D_s^{(*)+} D_s^{(*)-} X, \thinspace
D^{(*)}_s \mu \phi \nu X$ and
$c\bar{c} \rightarrow \mu \phi D^{(*)}_s$ are subtracted.
Using PDG~\cite{pdg} branching fractions in Eq.~\ref{d0ratio}, 
as well as combining the PDG measurement with 
BaBar's new measurement of 
$Br(D_s \rightarrow \phi \pi)$~\cite{babar_dsbr},
the preliminary branching fraction
\begin{eqnarray}
Br(B^0_s \rightarrow D_s^{(*)+} D_s^{(*)-}) = \nonumber \\ 
0.071 \pm 0.035 \thinspace {\mathrm{(stat.)}}
      ^{+0.029}_{-0.025} \thinspace {\mathrm{(syst.)}}
\end{eqnarray}
is measured.
Assuming that Eq.~\ref{dGsbr} is correct, 
\begin{equation}
\frac{\Delta\Gamma_{CP}}{\Gamma_s} \approx
\frac{\Delta\Gamma_s}{\Gamma_s} =
0.142 \pm 0.064 \thinspace {\mathrm{(stat.)}}
      ^{+0.058}_{-0.050} \thinspace {\mathrm{(syst.)}}
\end{equation}
is determined.

\subsection{Belle Limits on \boldmath$B^0_s$ Decay Modes}

In June 2005, a three-day engineering run at KEKB 
at the center-of-mass energy corresponding to the mass
of the $\Upsilon(5S)$ allowed Belle to collect 
1.86~fb$^{-1}$ of data~\cite{belleBs}, including produced
$B^{(*)}_s \bar{B}^{(*)}_s$ pairs.  From this
initial sample Belle was able to reconstruct 
the hadronic decays
$B^0_s \rightarrow D_s^+ \pi^-$, $K^+ K^-$,
and $D^{(*)+}_s D^{(*)-}_s$, the channel of interest.
Numbers of observed candidates for the latter were
too small to allow measurement of the branching fraction, 
but preliminary limits of:
\begin{eqnarray}
Br(B^0_s \rightarrow D_s^+ D_s^-) & < 7.1\%, \nonumber \\
Br(B^0_s \rightarrow D_s^{*+} D_s^-) & < 12.7\%, \nonumber \\
Br(B^0_s \rightarrow D_s^{*+} D_s^{*-}) & < 27.3\%,
\end{eqnarray}
were set.
KEKB is capable of producing 1~fb$^{-1}$ of integrated
luminosity per day at the $\Upsilon(5S)$, and a possible
50-day long run in the future can allow the measurement
of $Br(B^0_s \rightarrow D_s^{(*)+} D_s^{(*)-})$ to a
relative precision of 25\%, already competitive with
the Tevatron measurements.  Due to their lack of boost,
lifetime and oscillation measurements are significantly 
more difficult
at the $\Upsilon(5S)$, but the $B$ factories can contribute
to the measurement of these important branching fractions.

For completeness of reporting $B^0_s$ branching fraction
measurements, Belle has also taken advantage of their excellent
electromagnetic calorimeter to use this same data sample to
place a preliminary limit of 
$Br(B^0_s \rightarrow \gamma \gamma) < 0.56 \times 10^{-4}$
at 90\% C.L., already a factor of three improvement
over the PDG value~\cite{pdg}.
This decay mode is more sensitive to new physics than
other penguin decay modes, and $R$-parity violating
SUSY and fourth generation models can increase the 
SM prediction of
$(0.5 - 1.0) \times 10^{-6}$
for this branching fraction by up 
to two orders of magnitude.

\section{Combining and Comparing 
\boldmath$\Delta \Gamma_s$ Results}

\begin{figure}[htb]
\centering
\includegraphics[width=60mm]{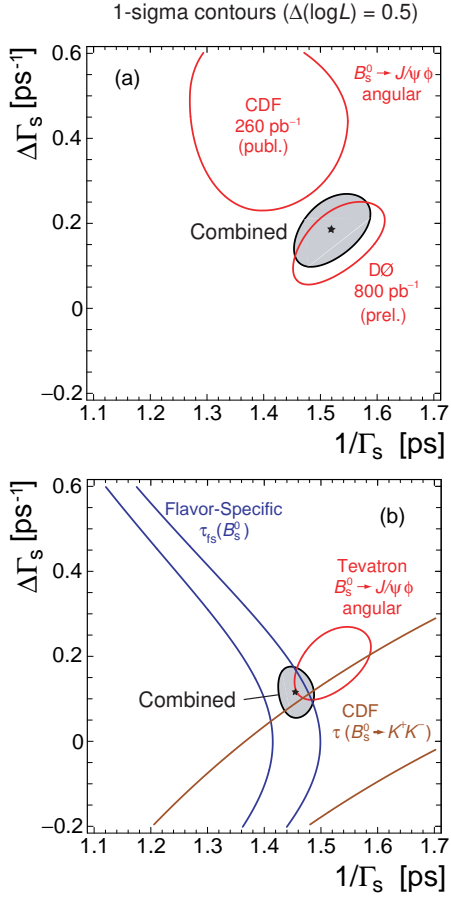}
\caption{
(a) Combination of the two experimental results
separating the $CP$-even and $CP$-odd components from 
angular distributions of products in the
decay $B^0_s \rightarrow J/\psi \phi$ as a function
of proper time;
(b) combination including the world average
flavor-specific $B^0_s$ lifetime (50\% $CP$-even, 
50\% $CP$-odd at time $t=0$ and the CDF result
of the $B^0_s$ lifetime from 
$B^0_s \rightarrow K^+K^-$ decays.
} \label{fig:jpsiphi}
\end{figure}

The most direct experimental results come from
the Tevatron from CDF~\cite{cdfjpsiphi} and
D\O~\cite{d0jpsiphi} (described elsewhere in
the proceedings~\cite{bslife}) where reconstructed
decays $B^0_s \rightarrow J/\psi \phi$ are separated
into $CP$-even and $CP$-odd components from fits to angular
distributions of $J/\psi$ and $\phi$ decay products
as a function of proper decay time.
A weighted average of CDF and D\O\ indicates that
this decay is $(17 \pm 4)\%$ $CP$ odd at time $t=0$, i.e., is
a dominantly $CP$-even decay.  Figure~\ref{fig:jpsiphi}(a)
shows the one sigma contours 
($\Delta \log{{\mathrm{(Likel.)}}} = 0.5$)
for the two experimental results and their 
combination (following the procedure outlined in Ref.~\cite{hfag}).
This combination results in a Tevatron average 
from $B^0_s \rightarrow J/\psi \phi$ decays 
of\footnote{The experimental measurements presented 
are really of $2 |\Gamma_{12}| \cos\phi$. The phase 
angle $\cos\phi$ is not measured, and in the context
of new physics, $\cos\phi$ can have any sign.}:
\begin{eqnarray}
\Delta\Gamma_s & = & 0.18 \pm 0.09 \thinspace
{\mathrm{ps}}^{-1}, \nonumber \\
\bar\tau = \frac{1}{\Gamma_s} & = & 1.520 \pm 0.068 \thinspace
{\mathrm{ps}}.
\end{eqnarray}

Precise measurements of the flavor-specific lifetime
of the $B^0_s$ can further constrain $\Delta \Gamma_s$
and $\Gamma_s$.  These flavor-specific decays are dominated by
semileptonic decays of the $B^0_s$, i.e.,
$B^0_s \rightarrow D_s \ell \nu$ where the flavor
of the meson, i.e., whether it is $B^0_s$ or $\bar{B}^0_s$
can be determined by the charge sign of the lepton.
These decays are 50\% $CP$-even and 50\% $CP$-odd at time
$t=0$, and each component decays away with a  different lifetime.
A superposition
of two exponentials thus results with decay
widths $\Gamma_s \pm \Delta \Gamma_s /2$.
Fitting to a single exponential results in a
measure of a flavor-specific
lifetime where~\cite{Hartkorn_Moser}:
\begin{equation}
\tau(B^0_s)_{\rm fs} = \frac{1}{\Gamma_s}
\frac{{1+\left(\frac{\Delta \Gamma_s}{2\Gamma_s}\right)^2}}
{{1-\left(\frac{\Delta \Gamma_s}{2\Gamma_s}\right)^2}
}.
\end{equation}
Updating the  world average flavor-specific lifetime value from the
Heavy Flavor Averaging Group
(HFAG)~\cite{hfag} with the new semileptonic
$B^0_s$ lifetime measurement from D\O\ submitted
for publication~\cite{d0bslife, bslife}, a value of
\begin{equation}
\tau(B^0_s)_{\rm fs} = 1.457 \pm 0.042 \thinspace 
{\mathrm{ps}}
\end{equation}
is obtained, giving the one-sigma band indicated on
Fig.~\ref{fig:jpsiphi}(b). 

Lastly, CDF has made a preliminary measurement~\cite{bskklife} 
of the $B^0_s$ lifetime
in decays $B^0_s \rightarrow K^+K^-$ of
\begin{equation}
\tau(B^0_s \rightarrow K^+K^-) = 1.53 \pm 0.18 \pm 0.02 \thinspace
{\mathrm{ps}}.
\label{theory_KK}
\end{equation}
This mode should be $CP$ even to within 5\%, and
hence measures the lifetime of the ``light" mass
eigenstate $\tau_L = 1/\Gamma_L$ and gives the one-sigma
constraint as indicated on Fig.~\ref{fig:jpsiphi}(b).

In the absence of new sources of $CP$ violation
in the penguin-dominated $b \rightarrow s u \bar{u}$ decay
amplitude, the lifetime in Eq.~\ref{theory_KK} corresponds
to $1/\Gamma_L$, the theory behind the
extraction of $\Delta \Gamma_s$ using these inputs is
considered to be valid and they are combined, to form
world averages of:
\begin{eqnarray}
\Delta\Gamma_s & = & 0.12 \pm 0.06 \thinspace 
{\mathrm{ps}}^{-1}, \nonumber \\
\bar\tau = \frac{1}{\Gamma_s} & = & 1.455 \pm 0.032 \thinspace
{\mathrm{ps}},
\end{eqnarray}
as shown by the shaded region in Fig.~\ref{fig:jpsiphi}.

\begin{figure}[htb]
\centering
\includegraphics[width=60mm]{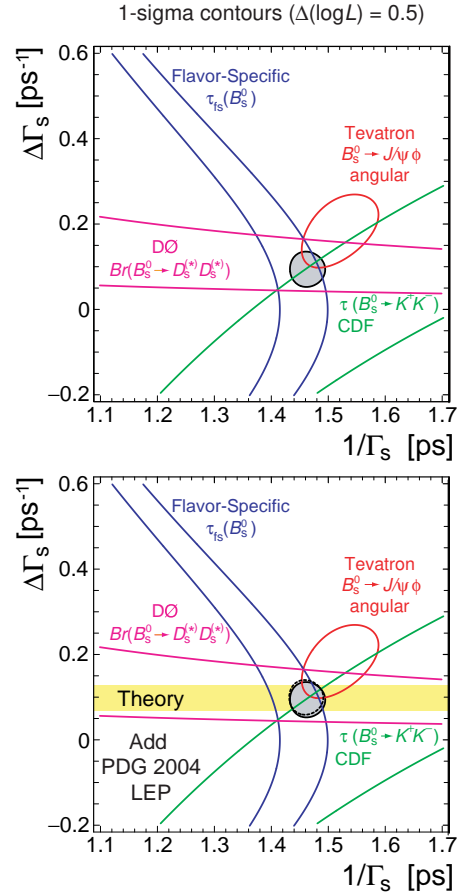}
\caption{
(a) Combination (shaded region) including the D\O\
measurement of 
$Br(B^0_s \rightarrow D^{(*)+}_s D^{(*)-})$;
(b) combination (dashed line) when all prior measurements
are included (PDG 2004), and comparison to the SM
prediction (horizontal band) with $f_{B_s}$ fixed
to 260~MeV.
} \label{fig:combine2}
\end{figure}

The previously described D\O\ measurement of 
$Br(B^0_s \rightarrow D^{(*)+}_s D^{(*)-})$ in 
conjunction with Eq.~\ref{dGsbr} can be
used to compare with this world
average.
Although there are worries that there may possibly be more
$CP$-odd component in this mode~\cite{Uli}, an additional
5\% theory systematic is added in quadrature and assuming
that Eq.~\ref{dGsbr} is valid, the resultant additional
constraint is shown in Fig.~\ref{fig:combine2}(a) and
combined with the previous inputs.

When all the inputs of PDG 2004~\cite{pdg2004} are
included, only a slight shift is observed in the total
world average (Fig.~\ref{fig:combine2}(b))
indicating the degree of weight of new
Tevatron measurements since 2004.
With all available inputs, the world average
values are then:
\begin{eqnarray}
\Delta\Gamma_s & = & 0.097 \pm 0.042 \thinspace
{\mathrm{ps}}^{-1}, \nonumber \\
\bar\tau = \frac{1}{\Gamma_s} & = & 1.461 \pm 0.030 \thinspace
{\mathrm{ps}}.
\end{eqnarray}
The result for $\Delta \Gamma_s$ is currently 2.3$\sigma$ from
zero.
A comparison to a theoretical prediction~\cite{nierste} of
\begin{equation}
\Delta \Gamma_s = (0.10 \pm 0.03) 
\left( \frac{f_{B_s}}
 {260 \thinspace {\mathrm{MeV}}} \right) ^2 \thinspace
 {\mathrm{ps}}^{-1},
\end{equation}
depicted as a horizontal shaded band in Fig.~\ref{fig:combine2}(b),
shows agreement with the SM, although errors are still large.
It should be noted that 
$\Delta \Gamma_s \approx 2 |\Gamma_{12}| \cos \phi$, and
new physics could result in  larger values of $\phi$  that would
then tend to {\em reduce} the measured value of
$\Delta \Gamma_s$.
The experimental result  can 
also be expressed as the two different
lifetimes of the mass eigenstates:
\begin{eqnarray}
\tau_L & = & \frac{1}{\Gamma_L} = 1.364 \pm 0.046 
\thinspace {\mathrm{ps}}, \nonumber \\
\tau_H & = & \frac{1}{\Gamma_H} = 1.573 \pm 0.061
\thinspace {\mathrm{ps}}.
\end{eqnarray}

Finally, we can test the predicted relationship~\cite{Bphysrun2}:
\begin{equation}
\frac{\Delta \Gamma_s}{\Delta m_s} \approx
\left| \frac{\Gamma_{12}}{M_{12}} \right|  =
\mathcal{O} \left( \frac{m_b^2}{M_W^2} \right) \approx
4 \times 10^{-3},
\end{equation}
with a more precise prediction~\cite{nierste} of:
\begin{equation}
\frac{\Delta \Gamma_s}{\Delta m_s} = (47 \pm 8) \times 10^{-4}.
\end{equation}
Using results on $\Delta m_s$~\cite{d0buchholz,bslife} and
the world average $\Delta \Gamma_s$ determined here, we find from
experimental measurements:
\begin{eqnarray}
\frac{\Delta \Gamma_s}{\Delta m_s} &  = & 
\frac{0.097 \pm 0.042 \thinspace {\mathrm{ps}^{-1}}}
{17.33 ^{+0.42}_{-0.21} \pm 0.07 \thinspace {\mathrm{ps}^{-1}}}
\nonumber \\
                                   & = &
				   (56 \pm 24) \times 10^{-4},
\end{eqnarray}
again showing (disappointingly!) agreement with
the SM prediction.

\section{Leptonic Decays of \boldmath$B$ Hadrons}

\subsection{Leptonic Decays 
\boldmath$B^0_d, B^0_s \rightarrow \mu^+ \mu^-$}

\begin{figure}[htb]
\centering
\includegraphics[width=70mm]{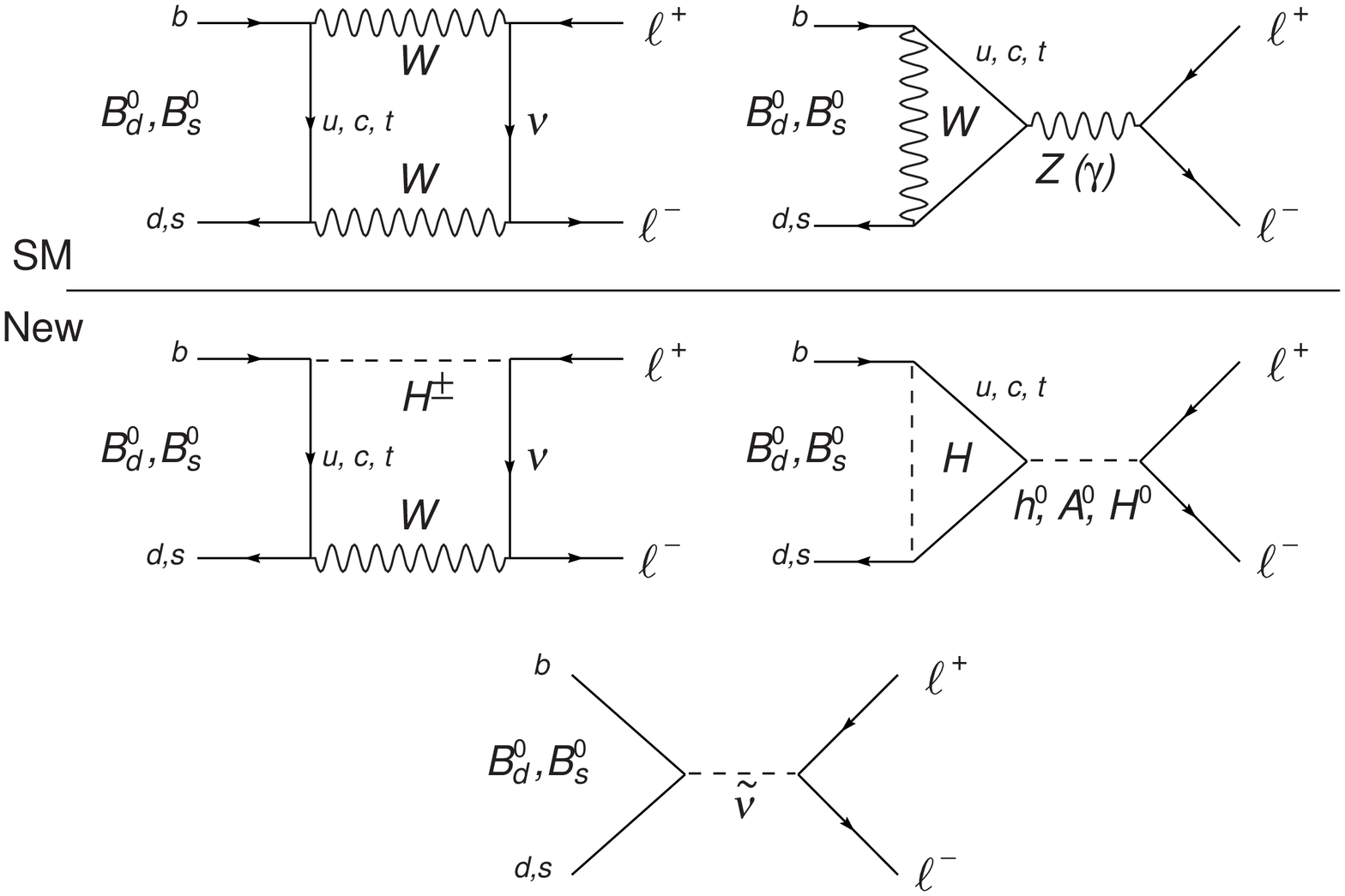}
\caption{
Feynman diagrams responsible for the decay
$B^0_d, B^0_s \rightarrow  \ell^+ \ell^-$.
} \label{fig:leptfigs}
\end{figure}

Decays of the type $B^0_d, B^0_s \rightarrow \mu^+ \mu^-$ provide
an excellent window into new physics that would tend
to enhance rates above their predicted SM values.
These decays 
represented by the diagrams of
Fig.~\ref{fig:leptfigs},
are highly suppressed by
a factor of $(m_{\ell}/m_B)^2$. As a result, decays into
electrons are effectively out of reach of collider
experiments, even if very much enhanced; decays into
$\tau$ leptons are the least suppressed, but it is difficult
to isolate these rare decays experimentally.
Decays into muons are therefore in the ``sweet spot", i.e., 
relatively straight forward to isolate experimentally, 
and within reach of observability if enhanced by new physics.
The SM predicts~\cite{buras}:
\begin{equation}
Br_{SM}(B^0_s \rightarrow \mu^+ \mu^-) = (3.35 \pm 0.32) \times 10^{-9},
\end{equation}
with $Br(B^0_d \rightarrow \mu^+ \mu^-)$ suppressed by
another factor of $|V_{td}/V_{ts}|^2 \approx 0.03$.

These branching fractions will grow as $\tan^6 \beta$ in the
Minimal Supersymmetric Standard Model (MSSM), and as 
$\tan^4 \beta$ in Two-Higgs Doublet Models (2HDM)
when charged Higgs bosons, $H^{\pm}$, take the place
of $W^{\pm}$ bosons in box and triangle diagrams, as
well as neutral SUSY Higgs bosons replacing $Z^0$ 
exchange~\cite{susy_mumu} as shown in Fig.~\ref{fig:leptfigs}
Contributions to enhanced signal are also possible 
through $s$-channel exchange of $R$-parity violating
SUSY particles.

\begin{figure}[htb]
\centering
\includegraphics[width=80mm]{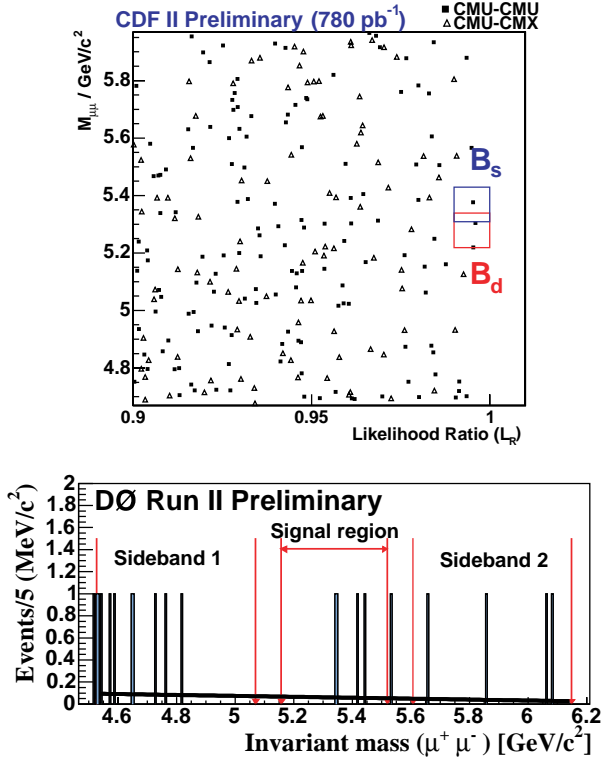}
\caption{
Top: CDF $\mu^+\mu^-$ invariant mass distribution versus the
event likelihood ratio (for $L_R > 0.90$) for events passing
baseline requirements; bottom: D\O\ $\mu^+\mu^-$ invariant 
mass distribution in signal and sideband regions.
} \label{fig:bsmumu}
\end{figure}

Limits on these decay rates are now dominated by the
Tevatron experiments.  Starting with dimuon triggers, and
requiring opposite-sign muons,
there are very large backgrounds due to
Drell-yan $\mu^+\mu^-$ continuum, sequential semimuonic
decays in $b \rightarrow c \rightarrow s$, 
double semileptonic decays $b \bar{b} \rightarrow 
\mu^+ \mu^- X$, and $b/c \rightarrow \mu^+ + $fake.
Both D\O~\cite{d0bsmumu} and CDF~\cite{cdfbsmumu,bskklife}
examine similar discriminating variables:
since the $B^0_s$ has lifetime, transverse
decay length significance or probability of 
significance; angle between the $\mu\mu$ vector and
decay length vector (i.e., the ``pointing consistency");
and isolation of the muons to reduce the background
due to multiple muons from regular $B$ hadron decays.
CDF combines these variables into a likelihood ratio, 
while D\O\ decides where to cut on each variable using
a random grid search for optimization.
After all cuts, the distribution
of the likelihood ratio versus invariant mass of the
$\mu^+ \mu^-$ pair for CDF,
and the $\mu^+\mu^-$  invariant mass for the D\O\ search is shown 
in Fig.~\ref{fig:bsmumu}.

Data sidebands are used to estimate expected backgrounds
in the search regions in each case, and limits derived from
the number of observed candidates compared to the number of expected.
Both experiments normalize the the number of mesons
observed in the $B^+ \rightarrow J/\psi (\rightarrow \mu^+\mu^-) K^+$
decay mode with the advantage of large statistics and
the muon identification efficiency being the same.
CDF has the mass resolution to separate the $B^0_d$ mass from
the $B^0_s$ mass, while D\O\ places limits on only $B^0_s$ decays.
Limit results are presented in Table~\ref{tab:bsmumu}.

\begin{table}[htb]
\begin{center}
\caption{Limits on the rare branching
fraction $Br(B^0_s \rightarrow \mu^+ \mu^-)$.}
\begin{tabular}{|c|c|c|c|}
\hline  \textbf{Experiment} & \textbf{Int. Lumin.} &
\textbf{Status} & \textbf{Limit (95\% C.L.)} 
\\
\hline 
CDF  & 780~pb$^{-1}$ & Prel. & $ < 1.0 \times 10^{-7} $ \\
CDF \& D\O & $\approx 300$~pb$^{-1}$  & Publ. & 
$ < 1.5 \times 10^{-7}$ \\[-4pt]
           &  each                   &       &  \\
D\O  & 700~pb$^{-1}$ & Prel. & $ < 2.3 \times 10^{-7} $ \\[-3pt]
      &               &       & (expected limit) \\
D\O  & 300~pb$^{-1}$ & Prel. & $ < 4.0 \times 10^{-7} $ \\
\hline
\end{tabular}
\label{tab:bsmumu}
\end{center}
\end{table}

The CDF preliminary limit on $Br(B^0_d \rightarrow \mu^+ \mu^-)$
is $< 3.0 \times 10^{-8}$, which is a factor approximately
three times more stringent than the next best limit
(from BaBar~\cite{babar_bsmumu}).
The CDF preliminary limit on $Br(B^0_s \rightarrow \mu^+ \mu^-)$ 
in the Table above is currently the world's best and
provides powerful constraints on new physics.
Examples are  stringent constraints 
on minimal SO(10) models with soft SUSY breaking~\cite{so10}, 
as well as mSUGRA models predicting neutralinos 
and cross sections consistent with relic 
density~\cite{dark_matter}, with 
branching fraction limits complementary to cross section limits
excluded by dark matter search experiments.

\subsection{Decay 
\boldmath$B^0_s \rightarrow \mu^+ \mu^- \phi$}

\begin{figure}[htb]
\centering
\includegraphics[width=70mm]{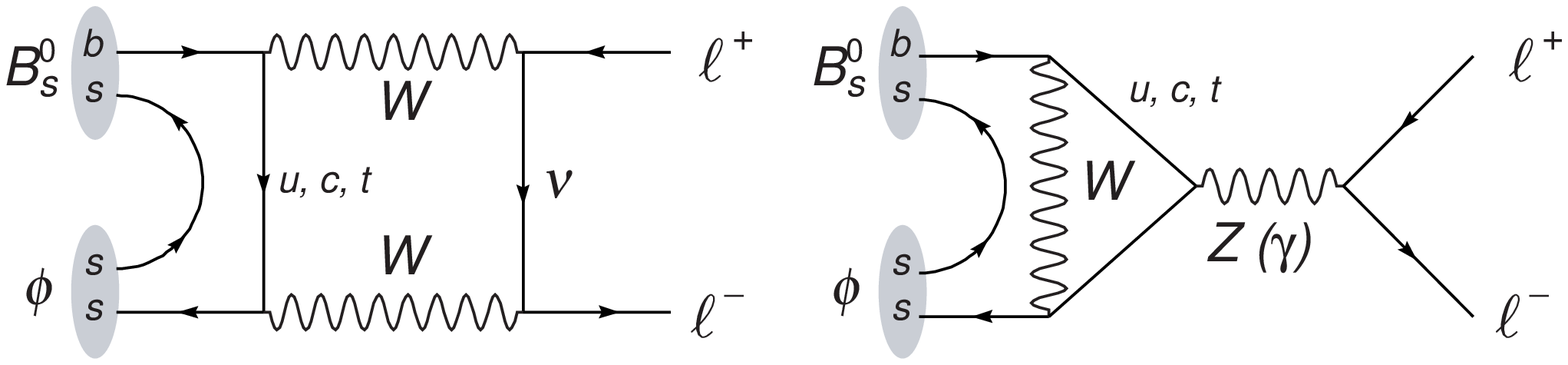}
\caption{
Feynman diagrams responsible for the decay
$B^0_s \rightarrow \phi \mu^+ \mu^-$.
} \label{fig:fcnc_diags}
\end{figure}

Although not strictly a purely leptonic decay, 
the  search for $B^0_s \rightarrow \mu^+ \mu^- \phi$ is reported
for completeness.  This flavor-changing neutral-current (FCNC)
decay can proceed through the diagrams of 
Fig.~\ref{fig:fcnc_diags} with a branching fraction
of $Br_{SM}(B^0_s \rightarrow \mu^+ \mu^- \phi) \approx 
1.6 \times 10^{-6}$ predicted by the SM, excluding
long-distance effects from charmonium resonances~\cite{bsmumuphi_the}.
The long term goal is to observe this mode to be
able to investigate FCNC $b \rightarrow s \ell^+ \ell^-$
transitions in the $B^0_s$ system, just as 
inclusive $B \rightarrow X_s \ell^+ \ell^-$ and
exclusive $B \rightarrow K^{(*)} \ell^+ \ell^-$ are
observed at $B$ factories and agree with the SM.
There is the potential for enhancement in these modes
in various SUSY and 2HDM models.

D\O\ has searched for the exclusive mode
$B^0_s \rightarrow \phi \mu^+ \mu^-$~\cite{d0bsmumuphi}.
The analysis follows the same procedure as for
searching for $B^0_s \rightarrow \mu^+ \mu-$ except that
a $\phi$ meson is added via the decay 
$\phi \rightarrow K^+ K^-$ and vertexed with the 
$\mu^+\mu^-$.  The same discriminating variables
are used, but the cut values are re-optimized.
Resonant regions of $J/\psi, \psi(2S) \rightarrow
\mu^+ \mu^-$ are removed, and normalization is made
to the resonant mode 
$B^0_s \rightarrow J/\psi (\rightarrow \mu^+ \mu^-) \phi$.
Zero events are observed in the signal region, with an
expectation of $1.6 \pm 0.4$ events.  From this a limit
of
\begin{equation}
Br(B^0_s \rightarrow \mu^+ \mu^- \phi) < 4.1 \times 10^{-6}
\thinspace (95\% \thinspace {\mathrm{C.L.}})
\end{equation}
is found.  This is a factor of ten improvement over 
the previous limit and only a factor of three more
than the SM prediction -- the Tevatron should be
able to observe this mode before the end of Run 2b.

\subsection{Leptonic Decay 
\boldmath$B^0_d \rightarrow \tau^+ \tau^-$}

The decay $B^0_d \rightarrow \tau^+ \tau^-$ is the least
helicity suppressed as pointed out previously, but is
the most difficult to isolate experimentally due to the
two to four missing neutrinos.  Limits previously did
not exist for this channel, and as a result, a ``loophole"
existed allowing for certain models involving 
leptoquark couplings and SUSY $\tan\beta$ enhancements above
the SM prediction of $Br_{SM}(B^0_d \rightarrow \tau^+\tau^-)
= 1.2 \times 10^{-7}$.

\begin{figure}[htb]
\centering
\includegraphics[width=80mm]{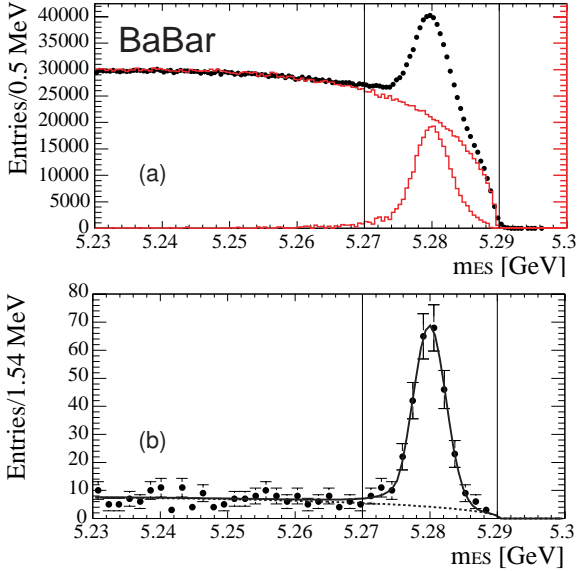}
\caption{
(a) Signal of 280k fully reconstructed $B^0_d$ 
companion-$B$ events in variable $m_{ES}$ (see text);
(b) remaining background events following all cuts to isolate 
$B^0_d \rightarrow \tau^+ \tau^-$ consistent with predicted
background.
} \label{fig:bdtautau}
\end{figure}

BaBar has now placed the first ever limit~\cite{babar_bdtautau}
on this channel. Starting with 280k fully reconstructed
$B^0_d \rightarrow D^{(*)} X$ decays (referred to as the 
companion $B$) as indicated by the peak in  
Fig.~\ref{fig:bdtautau}(a) in the variable
$m_{ES} = \sqrt{E^{*2}_{beam} - p^{*2}_B}$, where
$E^*_{beam}$ is the beam energy in the CM frame, and
$p^*_B$ is the reconstructed companion-$B$ momentum.
One-prong $\tau$ decays are then searched for in 
the rest of the event after removing
all neutral and charged kaons.  
Kinematics of charged daughter momenta and
the residual energy in the calorimeter are fed into  
an artificial neural network to separate signal from background.
After cutting on the network output, the signal
is as shown in in Fig.~\ref{fig:bdtautau}(b), with $263 \pm 19$
candidates observed in the data with an expected background
of $281 \pm 40$ candidates. From this, a limit:
\begin{equation}
Br(B^0_d \rightarrow \tau^+ \tau^-) < 3.4 \times 10^{-3}
\thinspace (90\% \thinspace {\mathrm{C.L.}})
\end{equation}
is determined.

\subsection{Leptonic Decay 
\boldmath$B^+ \rightarrow \tau^+ \nu_{\tau}$}

\begin{figure}[htb]
\centering
\includegraphics[width=30mm]{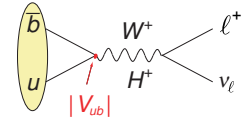}
\caption{
Feynman diagram responsible for 
$B^+ \rightarrow \tau^+ \nu_{\tau}$ decay in the SM, as well
as the possibility of probing for the existence of
charged Higgs bosons.
} \label{fig:taunu_diag}
\end{figure}

The decay $B^+ \rightarrow \ell^+ \nu_{\ell}$ proceeds through
a $W$-annihilation diagram as shown in Fig.~\ref{fig:taunu_diag} 
and the  branching fraction is then given by:
\begin{equation}
Br_{SM}(B^+ \rightarrow \ell^+ \nu_{\ell}) =
\frac{G^2_F m_B}{8\pi} m^2_{\ell}
(1 - \frac{m^2_{\ell}}{m^2_B})^2
f^2_B |V_{ub}|^2 \tau_B,
\end{equation}
providing access to the CKM matrix element $|V_{ub}|$, and
more importantly, to $f_B$ if $|V_{ub}|$ is taken as
an external input.
Alternatively, if a charged Higgs boson ($H^{\pm}$) exchange 
is possible, then
\begin{eqnarray}
Br(B^+ \rightarrow \ell^+ \nu_{\ell}) &  = &
  Br_{SM}(B^+ \rightarrow \ell^+ \nu_{\ell})\cdot \nonumber \\
 & & \cdot (1 - (\tan^2\beta/m^2_H)m^2_B)^2,
\end{eqnarray}
providing access to the SUSY parameter $\tan\beta$ and the
charged Higgs mass $m_H$~\cite{hou}.
There have been no new results in
$B^+ \rightarrow e^+ \nu_e$ or
$B^+ \rightarrow \mu^+ \nu_{\mu}$ since 2004, but there
is now the first observation of
$B^+ \rightarrow \tau^+ \nu_{\tau}$ from Belle.

\begin{figure}[htb]
\centering
\includegraphics[width=60mm]{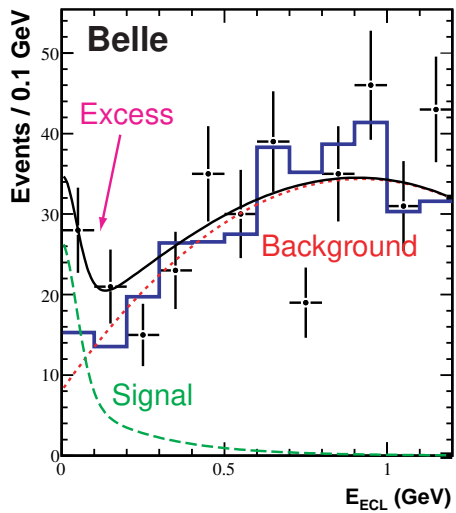}
\caption{
Distribution of excess energy in the electromagnetic
calorimeter after all selection requirements. Data
and background MC samples are represented by
points with error bars and the solid histogram,
respectively. The solid curve shows the result of the
fit with the sum of the signal shape (dashed) and
background shape (dotted).
} \label{fig:taunu}
\end{figure}

Belle's analysis~\cite{belle_taunu,belleBs}
proceeds in a similar fashion to the Babar analysis above, i.e.,
one $B^0_d$ is fully reconstructed, and in the remainder of the 
event, a search is made for a topology 
of one- or three-prong $\tau$ decays (in five
modes).  The properties of the remaining
event are compared with expected signal and background.
Figure~\ref{fig:taunu} shows the distribution of
excess energy in the electromagnetic calorimeter showing
a 4.0$\sigma$ excess over expectation at low values 
indicating the presence of  $B^+ \rightarrow \tau^+ \nu_{\tau}$
decays.
This excess translates into a measurement from
Belle of:
\begin{equation}
Br(B^+ \rightarrow \tau^+ \nu_{\tau}) =
(0.106 ^{+0.034 +0.018}_{-0.028 -0.016})\%,
\end{equation}
that is the first measurement of this 
branching fraction.  In the
the SM,
this can also be translated into a measurement of:
\begin{equation}
f_B |V_{ub}| = (7.73 ^{+1.24 +0.066}_{-1.02 -0.058})
\times 10^{-4} \thinspace {\mathrm{GeV}},
\end{equation}
and using the HFAG world average value of
$|V_{ub}|$~\cite{hfag},
\begin{equation}
f_B = 0.176 ^{+0.028 +0.020}_{-0.023 -0.018}
\thinspace {\mathrm{GeV}},
\end{equation}
that is the first direct measurement of this decay
parameter. Implications of this measurement for
constraints on the CKM unitarity triangle can
be found in Ref.~\cite{CKMfitter}.

For completeness, the BaBar limit~\cite{babar_taunu} on this 
branching fraction is
\begin{eqnarray}
Br(B^+ \rightarrow \tau^+ \nu_{\tau}) &  < & 2.6 \times 10^{-3}
\thinspace (90\% \thinspace {\mathrm{C.L.}}) \nonumber \\
                                     & = & (0.13 ^{+0.10}_{-0.09})\%.
\end{eqnarray}

\section{Summary}

In summary, the $B^0_s$ system is now being probed from all sides.
First evidence of $B^0_s$ oscillations and their frequency
gives $\Delta m_s$, but branching fraction measurements and
lifetimes into specific $CP$ eigenstate mixtures now gives a
world average of $\Delta \Gamma_s$ that is 2.3$\sigma$ from zero,
and also consistent with the SM.

In the field of leptonic and FCNC $B^0_s$ decays, the best limits
now come from the Tevatron, providing strong constraints on new physics
as they approach the SM predicted values.  These limits will improve
as Run 2b collects more data.  The first ever limit from BaBar on 
the branching fraction for $B^0_d \rightarrow \tau^+ \tau^-$ has
been presented, and first ever observation for the decay
$B^+ \rightarrow \tau^+ \nu_{\tau}$ has been shown by Belle.

\bigskip 
\begin{acknowledgments}
The author wishes to thank the organizers of the conference for
a very enjoyable and informative program, the many physics analysis
representatives
from D\O\, CDF, BaBar, and Belle for providing results, and
U. Nierste for useful discussions.
\end{acknowledgments}

\bigskip 

\end{document}